\documentclass[superscriptaddress,twocolumn,nofootinbib,showpacs,prl,amsmath]{revtex4}
\usepackage{times}
\usepackage{graphicx}
\usepackage{longtable}
\usepackage{color}
\usepackage{amsmath}
\usepackage{amssymb}

\newcommand{\beq}{\begin{equation}}
\newcommand{\eeq}{\end{equation}}
\newcommand{\beqa}{\begin{eqnarray}}
\newcommand{\eeqa}{\end{eqnarray}}
\newcommand{\ket} [1] {\vert#1\rangle}
\newcommand{\bra} [1] {\langle#1\vert}

\def\ket#1{|#1\rangle}
\def\bra#1{\langle #1|}

\def\opone{\leavevmode\hbox{\small1\kern-3.8pt\normalsize1}}

\begin{document}

\title{Experimental Entanglement and Nonlocality of a Two-Photon Six-Qubit Cluster State}
\author{Raino Ceccarelli}
\homepage{http://quantumoptics.phys.uniroma1.it/}
\affiliation{Dipartimento di Fisica della ``Sapienza''
Universit\`{a} di Roma,
Roma 00185, Italy and\\
Consorzio Nazionale Interuniversitario per le Scienze Fisiche della
Materia, Roma 00185, Italy}
\author{Giuseppe Vallone}
\homepage{http://quantumoptics.phys.uniroma1.it/}
\affiliation{Centro Studi e Ricerche ``Enrico Fermi'', Via Panisperna 89/A, Compendio del Viminale, Roma 00184, Italy}
\affiliation{Dipartimento di Fisica della ``Sapienza''
Universit\`{a} di Roma,
Roma 00185, Italy and\\
Consorzio Nazionale Interuniversitario per le Scienze Fisiche della
Materia, Roma 00185, Italy}
\author{Francesco De Martini}
\homepage{http://quantumoptics.phys.uniroma1.it/}
\affiliation{Dipartimento di Fisica della ``Sapienza''
Universit\`{a} di Roma,
Roma 00185, Italy and\\
Consorzio Nazionale Interuniversitario per le Scienze Fisiche della
Materia, Roma 00185, Italy} \affiliation{Accademia Nazionale dei
Lincei, via della Lungara 10, Roma 00165, Italy}
\author{Paolo Mataloni}
\homepage{http://quantumoptics.phys.uniroma1.it/}
\affiliation{Dipartimento di Fisica della ``Sapienza''
Universit\`{a} di Roma,
Roma 00185, Italy and\\
Consorzio Nazionale Interuniversitario per le Scienze Fisiche della
Materia, Roma 00185, Italy}
\author{Ad\'{a}n Cabello}
\affiliation{Departamento de F\'{\i}sica Aplicada II, Universidad de Sevilla, E-41012 Sevilla, Spain}
\date{\today}


\begin{abstract}
We create a six-qubit linear cluster state by transforming a
two-photon hyperentangled state in which three qubits are encoded
in each particle, one in the polarization and two in the linear momentum
degrees of freedom. ~For this state, we demonstrate genuine six-qubit
entanglement, persistency of entanglement against the loss of
qubits, and higher violation than in previous
experiments on Bell inequalities of the Mermin type.
\end{abstract}


\pacs{
03.67.Bg,
03.65.Ud,
42.50.Ex,
42.65.Lm
}

\maketitle


{\em Introduction.---}Progress in one-way quantum computing
\cite{raus01prl} requires the creation of $n$-qubit graph states
\cite{hein04pra} of high number of qubits. Graph states are also
fundamental resources for quantum nonlocality \cite{merm90prl,
guhn05prl, cabe07prl, cabe08pra}, quantum error correction
\cite{schl01pra}, and quantum entanglement \cite{hein04pra,
dur04prl}. In order to create multiqubit graph states it is
possible to increase the number of entangled particles
\cite{sack00nat, zhao03prl, walt05nat, walt05prl, kies05prl,
prev07nat, leib05nat, lu07nap} or to encode many qubits in each of
them \cite{vall07prl, chen07prl, vall08prl, gao08qph}. Multiqubit
graph states can be created by distributing the qubits between the
particles so that each particle carries one qubit. This is the way
in which four-qubit graph states with atoms \cite{sack00nat} and
photons \cite{zhao03prl, walt05nat, walt05prl, kies05prl,
prev07nat}, and six-qubit graph states with atoms \cite{leib05nat}
and photons \cite{lu07nap} were created. A second strategy is to
distribute the qubits so that each of the particles encodes two
qubits. This has been used to create two-photon four-qubit graph
states \cite{vall07prl, chen07prl, vall08prl} and up to five-photon
ten-qubit graph states \cite{gao08qph}. By generalizing this strategy, 
we have created a six-qubit two-photon linear cluster
state $|\widetilde {\rm LC}_6\rangle$, by
encoding three qubits in each particle: one qubit in the
polarization and two qubits in the linear momentum degrees of freedom 
(DOFs). The $|\widetilde {\rm LC}_6\rangle$ is the only
distribution of six qubits between two particles whose perfect
correlations have the same nonlocality as those of the six-qubit
Greenberger-Horne-Zeilinger (GHZ) state \cite{cabe08pra}, but only
requires two separated carriers \cite{cabe07prl}.




\begin{figure}[b]
\begin{center}
\includegraphics[width=0.8\columnwidth]{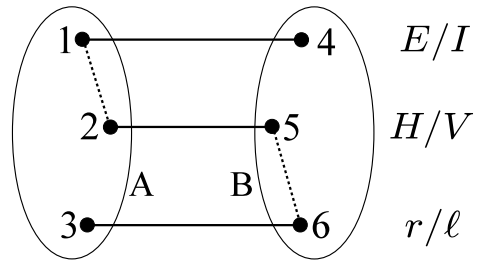}
\caption{Graph associated to a two-photon six-qubit entangled state.
Each set represents a photon and each vertex corresponds to a qubit.
Each link represents a CZ operation between the two connected
qubits. Dashed lines represent links present in the
$\ket{\text{LC}_{6}}$ state and absent in the $\ket{\rm HE_6}$
state. In the experiment, qubits 1 and 4 are encoded into
external/internal ($E$/$I$) modes, qubits 2 and 5 into
horizontal/vertical ($H$/$V$) polarization, and qubits 3 and 6 into
right/left ($r$/$\ell$) modes. See the text for
details.}\label{fig:graph}
\end{center}
\end{figure}


Consider the graph in Fig.~\ref{fig:graph} and associate a single
qubit to each vertex. The linear cluster state $|{\rm LC}_6\rangle$
is defined as the only six-qubit state which satisfies $g_i|{\rm
LC}_6\rangle=|{\rm LC}_6\rangle$, $\forall i$, where $g_i$
corresponds to the vertex $i$ of the graph in Fig.~\ref{fig:graph},
and is defined as $g_i= X_i \bigotimes_{j \in {\cal N}(i)} Z_j$,
where, e.g., $X_i$ is the Pauli matrix $\sigma_x$ of qubit $i$, and
${\cal N}(i)$ is the set of vertices which are connected to $i$. An
equivalent definition of graph states can be given in terms of
Controlled-$Z$ operations defined on qubits $i$ and $j$ as
CZ$_{ij}={\ket0}_i\bra0\otimes\openone_j+{\ket1}_i\bra1\otimes Z_j$.
The graph state $\ket{\mathcal G}$ associated to the $N$-vertex
graph $\mathcal G$ can be written as
\begin{equation}
\ket{\mathcal G}=\Bigl(\prod_{\langle i,j\rangle}{\rm
CZ}_{ij}\Bigr)\bigotimes^N_{i=1}\ket{+}_i, \label{eq:graph}
\end{equation}
where ${\langle i,j\rangle}$ indicates the connected vertices in
$\mathcal G$ and $\ket+_i=\frac{1}{\sqrt2}(\ket0_i+\ket1_i)$.

The specific distribution of the six qubits between the two photons
in Fig.~\ref{fig:graph} (qubits 1, 2, and 3 are carried by photon
$A$, and qubits 4, 5, and 6 by photon $B$) allows bipartite
nonlocality \cite{cabe07prl} because, in this distribution, all the
single-qubit Pauli observables satisfy EPR's criterion for elements
of reality \cite{eins35pr}, since the result of measuring any of the
Pauli observables on qubits 1, 2, and 3 can be
predicted with certainty from measurements on qubits 4, 5, and 6, 
and viceversa. This property is not satisfied
by other methods of creating graph states using different DOFs 
of the same photon, where new qubits are added by local
operations \cite{gao08qph}.






\begin{figure*}[t]
\begin{center}
\centering\includegraphics[width=1.46\columnwidth]{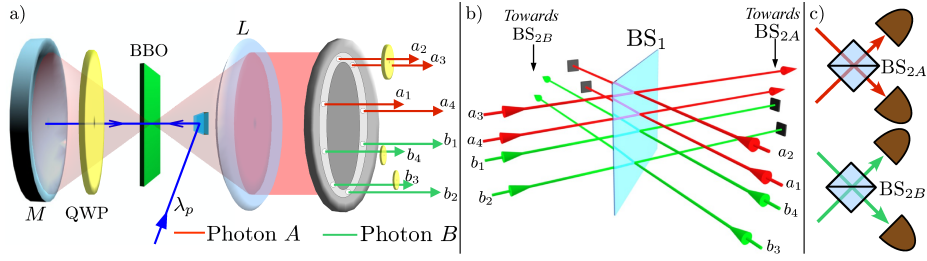}
\caption{Generation of the six-qubit linear cluster state. a) Scheme
of the entangled two-photon six-qubit parametric source: a UV laser
beam (wavelength $\lambda_p$) impinges on the Type I
BBO crystal after reflection on a small mirror. The
polarization entangled state $(|H\rangle|H\rangle -
|V\rangle|V\rangle)/\sqrt{2}$ arises from the superposition of the
degenerate emission cones of the crystal. Basic elements of
the source are: i) a spherical mirror ($M$), reflecting both the
parametric radiation and the pump beam, whose micrometric
displacement enables phase control between the
$\ket{H}\ket{H}$ and $\ket{V}\ket{V}$ events. ii) A quarter
wave-plate (WP), located between mirror $M$ and BBO, performing the
$|H\rangle|H\rangle \rightarrow |V\rangle|V\rangle$ transformation
on the left cone. iii) A positive lens $L$, transforming the conical
parametric emission of the crystal into a cylindrical one. Four
pairs of correlated longitudinal modes $a_i$-$b_i$ ($i=1,\ldots,4$)
are selected by an eight-hole screen. One half WP oriented at
$45^\circ$ intercepting modes $a_2$, $a_3$ and two half WPs
oriented at $0^\circ$ intercepting $b_3$, $b_4$ determine the
transformation from $|{\rm \widetilde{HE}}_6\rangle$ to $|\widetilde{\rm
LC}_6\rangle$. b) Spatial superposition between the left ($\ell$) and right ($r$)
modes on the common 50/50 beam splitter BS$_1$. $\ell$ modes $a_1$,
$a_2$ ($b_3$, $b_4$) are respectively matched with $r$ modes
$a_4$, $a_3$ ($b_2$, $b_1$) on the $A$ ($B$) side. Temporal
indistinguishability is obtained by setting to zero the path delay between
the right and the left modes. c) Spatial superposition
between the internal $I$ ($a_2$, $a_3$, $b_2$, $b_3$) and external $E$ ($a_1$, $a_4$, $b_1$, $b_4$)
modes is performed on BS$_{2A}$ and
BS$_{2B}$ for the $A$ and $B$ photon, respectively. Independent
adjustment of time delay between the $A$ and $B$ modes coming out of
BS$_1$ determines interference between the modes. 
After BS$_1$, only the $A$ ($B$) modes contribute to the interference 
on BS$_{2A}$ (BS$_{2B}$), while the others modes are intercepted by beam stops.}
\label{fig:setup}
\end{center}
\end{figure*}

{\em Experimental preparation.---}We create the state
$|\widetilde{\rm LC}_6\rangle$, equivalent up to single qubit
unitary transformations to $|{\rm LC}_6\rangle$, in two steps:
first, we prepare a six-qubit hyperentangled state ($|{\rm \widetilde{HE}}_6\rangle$) [cf. Fig.~\ref{fig:graph}] 
by a triple entanglement of two photons. 
The quantum information is encoded in the
polarization (qubits 2 and 5) and longitudinal momentum (qubits 1
and 4, and 3 and 6) photon DOFs. Then, we
transform $|{\rm \widetilde{HE}}_6\rangle$ into $|\widetilde{\rm LC}_6\rangle$
by applying a sequence of unitary transformations which entangle
qubits 1 and 2, and qubits 5 and 6.

The experimental setup used to create and measure the $|\widetilde
{\rm LC}_6\rangle$ is illustrated in Fig.~\ref{fig:setup}.
We used spontaneous parametric down-conversion (SPDC) in a single 0.5 mm thick Type I
$\beta$-barium- borate (BBO) crystal excited by a continuous wave
UV laser, following a scheme described in Fig.~\ref{fig:setup} \cite{barb05pra,vall09pra}. 
Precisely, four pairs of correlated spatial modes \cite{ross09prl}, labeled as $\ell$ ($r$) for the left
(right) side of the emission cone and as $I$ ($E$) considering the
internal (external) modes [cf. Fig.~\ref{fig:setup}(a)] were selected within the conical emission of the crystal. 
The starting point for the cluster state generation was the six-qubit
$|{\rm \widetilde{HE}}_6\rangle$, given by the product of one polarization 
and two longitudinal momentum entangled states,
\begin{align}
&\ket{\widetilde{\rm HE}_6} = 
\frac1{\sqrt2}\left({\ket{EE}}_{AB}+{\ket{II}}_{AB}\right)\label{eq:HE}
\\
&\otimes\frac1{\sqrt2}\left({\ket{HH}}_{AB}-{\ket{VV}}_{AB}\right)
 \otimes\frac1{\sqrt2}\left({\ket{\ell
r}}_{AB}+{\ket{r\ell}}_{AB}\right), \nonumber
\end{align}
where $A$ ($B$) corresponds to the up (down) side of the
conical crystal emission.

By using the following correspondence
between physical states and qubit states
\begin{subequations}\label{eq:correspondence}
\begin{align}
\{|E\rangle_A, |I\rangle_A\} \rightarrow \{|0\rangle_1, |1\rangle_1\}\,, \\
\{|H\rangle_A, |V\rangle_A\} \rightarrow \{|0\rangle_2, |1\rangle_2\}\,, \\
\{|r\rangle_A, |\ell\rangle_A\} \rightarrow \{|0\rangle_3, |1\rangle_3\}\,, \\
\{|E\rangle_B, |I\rangle_B\} \rightarrow \{|0\rangle_4, |1\rangle_4\}\,, \\
\{|H\rangle_B, |V\rangle_B\} \rightarrow \{|0\rangle_5, |1\rangle_5\}\,, \\
\{|r\rangle_B, |\ell\rangle_B\} \rightarrow \{|0\rangle_6, |1\rangle_6\}\,,
\end{align}
\end{subequations}
the hyperentangled state \eqref{eq:HE} is equivalent, up to single
qubit unitary transformations, to the graph state $\ket{\rm HE_6}$
shown in Fig.~\ref{fig:graph}. Specifically, $\ket{{\rm \widetilde{HE}}_6}=H_2X_3H_3H_4Z_5\ket{{\rm HE}_6}$, where $H_i$ denotes the
Hadamard operation on qubit $i$. By Eq. \eqref{eq:graph}, the
cluster state $\ket{{\rm LC}_6}$ is obtained from $\ket{\rm HE_6}$ by
applying the CZ$_{12}$ and CZ$_{65}$ gates. Then, by applying the
gates CX$_{12}$ (a Controlled-$X$ operation) and CZ$_{65}$ on the
hyperentangled state $\ket{{\rm \widetilde{HE}}_6}$, we obtain
\begin{equation}\label{eq:unitaries}
\ket{\widetilde{\rm LC}_6}={\rm CX}_{12}{\rm CZ}_{65}\ket{{\rm \widetilde{HE}}_6}=H_2X_3H_3H_4Z_5\ket{{\rm LC}_6}.
\end{equation}

The created state, $|\widetilde{{\rm LC}}_6 \rangle$, is, up to a
unitary transformation, equivalent to the two-photon six-qubit
cluster state $|{\rm LC}_6 \rangle$ by the correspondence
\eqref{eq:unitaries}. Specifically, the relation given in
\eqref{eq:unitaries} between $| \widetilde{{\rm LC}}_6 \rangle$ and
$|{\rm LC}_6 \rangle, $ implies that $|\widetilde{{\rm LC}}_6
\rangle$ is the only common eigenstate of the generators
$\widetilde{g}_i$ obtained from $g_i$ by changing $X_2
\leftrightarrow Z_2$, $X_3 \rightarrow -Z_3$, $Z_3 \rightarrow X_3$,
$X_4 \leftrightarrow Z_4$, and $X_5 \rightarrow -X_5$. Qubits 1 and 4
are encoded by the $E/I$ degree of freedom, qubits 2 and 5 by the
$H/V$ polarization, and qubits 3 and 6 by the $r/\ell$ degree of
freedom.

The $|\widetilde{\rm LC}_6\rangle$ state can be written as
\begin{equation}
\begin{aligned}
\ket{\widetilde {\rm LC}_6}&=\frac1{2}\left[{\ket{EE}}_{AB}{\ket{\phi^-}}_{\pi}{\ket{\ell r}}_{AB}
+{\ket{EE}}_{AB}{\ket{\phi^+}}_{\pi}{\ket{r\ell}}_{AB}\right.\\
&\left.-{\ket{II}}_{AB}{\ket{\psi^-}}_{\pi}{\ket{\ell r}}_{AB}+
{\ket{II}}_{AB}{\ket{\psi^+}}_{\pi}{\ket{r\ell}}_{AB}\right],
\end{aligned}
\end{equation} where
$\ket{\phi^{\pm}}_\pi=\frac{1}{\sqrt2}({\ket{HH}}_{AB}\pm{\ket{VV}}_{AB})$
and
$\ket{\psi^{\pm}}_\pi=\frac{1}{\sqrt2}({\ket{HV}}_{AB}\pm{\ket{VH}}_{AB})$
are the standard polarization Bell states.

The transformation from the hyperentangled state to the cluster state 
was carried out by two wave-plates intercepting the $|{\rm \widetilde{HE}}_6\rangle$'s
output modes. Precisely, since qubits 1 ($E/I$) and 2 ($H/V$) are
encoded in photon $A$, the CX$_{12}$ gate was obtained by applying a
half wave-plate (WP) oriented at $45^\circ$ on the internal $A$ modes
($a_2$ and $a_3$ in Fig.~\ref{fig:setup}). Equivalently, the
CZ$_{65}$ was obtained by inserting a half WP oriented at
$0^\circ$ on the left $B$ modes ($b_3$ and $b_4$). In the actual
experiment, we used one WP intercepting both $a_2$ and $a_3$
modes, while one WP was used for the $b_3$ mode and one
for the $b_4$ mode [see Fig.~\ref{fig:setup}a)].

The experimental setup sketched in Fig.~\ref{fig:setup}b) and
\ref{fig:setup}c) allows the simultaneous measurement of three single
qubit compatible observables for each particle. It is given by two
chained interferometers whose core elements are given by three
symmetric (50/50) beam splitters BS$_1$, BS$_{2A}$, and BS$_{2B}$.
In BS$_1$, the four $\ell$ modes are made indistinguishable from the
corresponding $r$ modes both in space and time, while $I$ and $E$
modes belonging to the $A$ ($B$) side are matched on BS$_{2A}$
(BS$_{2B}$). Two pairs of single photon detectors detect the
output modes $A$ or $B$, while polarization entanglement is 
measured by four polarization analyzers (not shown in the Figure), 
one for each detector. Nearly 500 coincidences per second were detected,
which is 4 orders of magnitude larger than the rate of the
six-photon linear cluster state \cite{lu07nap}.




{\em Fidelity.---}We measured the fidelity of our preparation by
measuring the $64$ stabilizers $\widetilde{s}_i$ of the
$|\widetilde{{\rm LC}}_6 \rangle$, i.e., all the products of the
generators $\widetilde{g}_i$. We obtained (see Table I)
\begin{equation}
F= \frac{1}{64} \sum_{i=1}^{64} \langle\widetilde{s}_i\rangle = 0.6350 \pm 0.0008,
\end{equation}
which constitutes an improvement of 7\% with respect to the best
previous fidelity for six-qubit graph states with six particles
\cite{leib05nat, lu07nap}. The fidelity value is limited by imperfections in phase and
polarization settings, such as the two controlled operations (CX and
CZ), and mainly by non perfect mode matching on the three beam splitters (BSs). 
Note that the measurements on the second momentum (I/E qubit) are naturally affected by imperfections of the first momentum setup.
Using single mode fibers combined with integrated quantum optical circuits
in the experimental setup would allow to largely restore the state
fidelity \cite{poli08sci}. Other DOFs, such as time-energy and orbital 
angular momentum, could be adopted to increase the number of qubits.
However, this imposes the use of optical components of high quality to 
preserve the fidelity.




{\em Entanglement witness.---}We tested whether or not the created
state has genuine six-qubit entanglement (i.e., inexplicable by five
or less qubit entanglement). For that purpose, we measured an
entanglement witness specifically designed \cite{toth05prl} to
detect genuine six-qubit entanglement around the $|\widetilde{{\rm
LC}}_6\rangle$,
\begin{equation}
\mathcal W_F = \openone - 2\ket{\widetilde{\rm LC}_6}\bra{\widetilde{\rm LC}_6}=
\openone -\frac{1}{32} \sum_{i=1}^{64}\widetilde{s}_i,
\end{equation}
where $\openone$ is the identity operator. There is entanglement
whenever
\begin{equation}
\langle\mathcal W_F\rangle =1-2F < 0.
\end{equation}
We obtained,
\begin{equation}
\mathcal W_{F} = -0.270 \pm 0.002,
\end{equation}
which is negative by 135 standard deviations and thus proves the
existence of a genuine six-qubit entanglement.




{\em Quantum nonlocality.---}The specific state we have created is
the only distribution of six qubits between two particles whose
perfect correlations have the same nonlocality as those of the
six-qubit Greenberger-Horne-Zeilinger (GHZ) state \cite{cabe08pra}
and, instead of requiring six separated carriers to show
nonlocality, it only requires two \cite{cabe07prl}. In any local
theory in which all the single-qubit Pauli observables can be
regarded as elements of reality in the sense of EPR \cite{eins35pr},
the following Bell inequality \cite{cabe08pra} must hold:
\begin{equation}
\mathcal B \le 4\equiv B_{\text {LHV}}, \label{idelineq}
\end{equation}
where
\begin{equation}
\mathcal B = |\widetilde{g}_1 (\openone+\widetilde{g}_2)
(\openone+\widetilde{g}_3) (\openone+\widetilde{g}_4)
(\openone+\widetilde{g}_5) \widetilde{g}_6|.
\end{equation}
This inequality is the optimal one to detect nonlocality even when
the $|\widetilde{{\rm LC}}_6\rangle$ has a maximum amount of white
noise \cite{cabe08pra}. EPR's assumption is that single-qubit observables on photon A (B)
are elements of reality (i.e., have pre-assigned outcomes) when
their outcomes can be predicted with probability 1 from measurements
on photon B (A). However, in our experiment, the single-qubit
observables on photon A (B) in the inequality \eqref{idelineq} can be
predicted from measurements on photon B (A) with probabilities
ranging from 0.78 to 0.94. Therefore, we need to relax EPR's
assumption and assume that single-qubit Pauli observables are elements of reality if
they can be predicted with probability higher than 0.77. For example, 
if $\langle X_{3}Z_{5} X_{6}\rangle= 1- \epsilon$, with $0 \le \epsilon \ll 1$, then
a fraction $\epsilon$ ($1-\epsilon$) of the pairs are uncorrelated (perfectly correlated). 
Therefore, the outcome of $X_3$ in photon $A$ can be
correctly predicted from the outcome of $Z_5 X_6$ in photon $B$ with
probability $1$ for the the correlated pairs and with probability $\frac12$ for the uncorrelated pairs.
Thus the outcome of $X_3$ can be predicted with probability $1(1-\epsilon)+\frac12\epsilon=1-\frac\epsilon2$.

We tested the Bell inequality
(\ref{idelineq}) and obtained
\begin{equation}\label{Bexp}
\mathcal B_{\rm exp} = 7.018 \pm 0.028,
\end{equation}
equivalent to a degree of nonlocality $\mathcal D=\frac{\mathcal
B_{\rm exp}}{B_{\text{LHV}}}$ of $1.7545 \pm 0.0070$, which is a
considerable improvement compared to previous violations of Bell
inequalities only involving perfect correlations and using
four-qubit states: $(2.59 \pm 0.08)/2=1.29$ \cite{walt05prl}, $(2.73
\pm 0.12)/2=1.36$ \cite{kies05prl}, $(3.4145 \pm 0.0095)/2=1.70$
\cite{vall07prl} and $(2.50 \pm 0.04)/2=1.25$ \cite{zhou08pra}. A
higher value of $\cal D$ has been reached for a Bell inequality not
involving perfect correlations \cite{zhao03prl, arde92pra, guhn08pra}. 
To our knowledge, the result of Eq. \eqref{Bexp} represents the first nonlocality test
with a six-qubit graph state. The fact that we have obtained a higher
degree of nonlocality than with simpler systems is an experimental
confirmation that quantum nonlocality can increase as the complexity
of the system grows in spite of the decrease of the fidelity \cite{merm90prl}.




{\em Persistency of entanglement.---}Linear cluster states are
particular entangled states that, when some qubits are lost, still
present some entanglement and nonlocality \cite{raus01prl}. Here we
can check that, by tracing qubits 3 and 6 or, alternatively, qubits
1 and 4, the remaining four qubits are still entangled and violate a
Bell inequality. Indeed, we observed the violation of the two following
Bell inequalities:
\begin{subequations}
\begin{align}
&\beta \le 2, \label{idelineqa} \\
&\beta' \le 2, \label{idelineqb}
\end{align}
\end{subequations}
where,
\begin{subequations}
\begin{align}
&\beta = |\widetilde{g}_1 (\openone+\widetilde{g}_2) (\openone+\widetilde{g}_4)|, \\
&\beta' = |(\openone+\widetilde{g}_3) (\openone+\widetilde{g}_5)
\widetilde{g}_6|.
\end{align}
\end{subequations}
The first is a 2-2-0-2-1-0-setting Bell inequality (i.e., it only
involves measurements on qubits 1, 2, 4, and 5); the second is a
0-1-2-0-2-2-setting Bell inequality (i.e., it only involves
measurements on qubits 2, 3, 5, and 6). We tested these two Bell
inequalities, obtaining
\begin{subequations}
\begin{align}
&\beta_{\rm exp} = 2.325 \pm 0.014,\\
&\beta'_{\rm exp} = 2.881 \pm 0.012.
\end{align}
\end{subequations}
These results correspond, respectively, to a violation of 23 and 73
standard deviations.
The fact that the violation in (15a) is lower than that in (15b) is
due to the critical E/I mode matching occurring on BS$_{2A}$, and BS$_{2B}$.
We attribute this to the angular divergence of the selected modes that 
enhances their transverse size in the measurement setup.




{\em Conclusions.---}In this Letter we have presented the first
experimental demonstration of a six-qubit linear cluster state built
on a two-photon triple entangled state. An entanglement witness has
been measured for this state and its persistency of entanglement and
quantum nonlocality properties have been characterized in detail.
Cluster states built on two photons and more DOFs
present both advantages and disadvantages with respect to multi-photon
cluster states. On one side, no more than few pairs of photons at a
time are created by SPDC, due to the probabilistic nature of this
process; then, multi-photon detection is seriously affected by the
limited quantum efficiencies of detectors; finally, an entangled
state built on a large number of particles is more affected by
decoherence because of the increased difficulty of making photons
indistinguishable. On the other side, increasing the number of
DOFs implies an exponential requirement of resources,
for instance, $2^{N}$ $\mathbf{k}$-modes per photon must be selected within the
emission cone to encode {N} qubits in each photon. Despite that,
working with a limited number of DOFs (up to four) is
still more convenient than increasing the number of photon pairs.
Hence a hybrid approach (i.e., multi-DOF/multi-photon states) can be
conceived in a medium-term time scale to overcome the structural
limitations in generation/detection of quantum photon states.




The authors thank O. G\"{u}hne for useful conversations. This work
was supported by the Spanish MCI Project No. FIS2008-05596 and the
Junta de Andaluc\'{\i}a Excellence Project No. P06-FQM-02243 and by
Finanziamento Ateneo 08 Sapienza Universit\'{a} di Roma.\\
\vskip-1cm

\begin{table*}[h]
\caption{Experimental results: measurement of the
64 stabilizers $\widetilde s_i$ of $\ket{\widetilde{\text{LC}}_6}$, i.e., all the products
of the generators $\widetilde g_i$. Last three columns indicate in which Bell inequality test
each experimental value was used.
}
\begin{tabular}[c]{|ccccc|}
\hline \hline
Stabilizer & Experimental value & $\mathcal{B}_{\mathrm{exp}}$ & $\beta$ & $\beta^{\prime}$ \\
\hline
$1$ & $1.0000 \pm 0.0000$ &  &  &  \\
$\tilde g_1$ & $0.5928 \pm 0.0075$ &  & \checkmark &  \\
$\tilde g_2$ & $0.8788 \pm 0.0053$ &  &  &  \\
$\tilde g_3$ & $0.9984 \pm 0.0005$ &  &  &  \\
$\tilde g_4$ & $0.9973 \pm 0.0008$ &  &  &  \\
$\tilde g_5$ & $0.7905 \pm 0.0057$ &  &  &  \\
$\tilde g_6$ & $0.8310 \pm 0.0062$ &  &  & \checkmark \\
$\tilde g_1 \tilde g_2$ & $0.5657 \pm 0.0059$ &  & \checkmark &  \\
$\tilde g_1 \tilde g_3$ & $0.5930 \pm 0.0075$ &  &  &  \\
$\tilde g_1 \tilde g_4$ & $0.5602 \pm 0.0076$ &  & \checkmark &  \\
$\tilde g_1 \tilde g_5$ & $0.5872 \pm 0.0076$ &  &  &  \\
$\tilde g_1 \tilde g_6$ & $0.4653 \pm 0.0095$ & \checkmark &  &  \\
$\tilde g_2 \tilde g_3$ & $0.8586 \pm 0.0062$ &  &  &  \\
$\tilde g_2 \tilde g_4$ & $0.8775 \pm 0.0053$ &  &  &  \\
$\tilde g_2 \tilde g_5$ & $0.7042 \pm 00066$ &  &  &  \\
$\tilde g_2 \tilde g_6$ & $0.8288 \pm 0.0062$ &  &  &  \\
$\tilde g_3 \tilde g_4$ & $0.9970 \pm 0.0009$ &  &  &  \\
$\tilde g_3 \tilde g_5$ & $0.7896 \pm 0.0057$ &  &  &  \\
$\tilde g_3 \tilde g_6$ & $0.7484 \pm 0.0056$ &  &  & \checkmark \\
$\tilde g_4 \tilde g_5$ & $0.7339 \pm 0.0084$ &  &  &  \\
$\tilde g_4 \tilde g_6$ & $0.8312 \pm 0.0062$ &  &  &  \\
$\tilde g_5 \tilde g_6$ & $0.6392 \pm 0.0060$ &  &  & \checkmark \\
$\tilde g_1 \tilde g_2 \tilde g_3$ & $0.4504 \pm 0.0092$ &  &  &  \\
$\tilde g_1 \tilde g_2 \tilde g_4$ & $0.6063 \pm 0.0074$ &  & \checkmark &  \\
$\tilde g_1 \tilde g_2 \tilde g_5$ & $0.5378 \pm 0.0086$ &  &  &  \\
$\tilde g_1 \tilde g_2 \tilde g_6$ & $0.4169 \pm 0.0065$ & \checkmark &  &  \\
$\tilde g_1 \tilde g_3 \tilde g_4$ & $0.5603 \pm 0.0076$ &  &  &  \\
$\tilde g_1 \tilde g_3 \tilde g_5$ & $0.5874 \pm 0.0075$ &  &  &  \\
$\tilde g_1 \tilde g_3 \tilde g_6$ & $0.4651 \pm 0.0063$ & \checkmark &  &  \\
$\tilde g_1 \tilde g_4 \tilde g_5$ & $0.5882 \pm 0.0074$ &  &  &  \\
$\tilde g_1 \tilde g_4 \tilde g_6$ & $0.4148 \pm 0.0075$ & \checkmark &  &  \\
$\tilde g_1 \tilde g_5 \tilde g_6$ & $0.4450 \pm 0.0061$ & \checkmark &  &  \\
 \hline \hline
\end{tabular}
 \qquad
\begin{tabular}[c]{|ccccc|}
\hline \hline
Stabilizer & Experimental value & $\mathcal{B}_{\mathrm{exp}}$ & $\beta$ & $\beta^{\prime}$ \\
\hline
$\tilde g_2 \tilde g_3 \tilde g_4$ & $0.8592 \pm 0.0062$ &  &  &  \\
$\tilde g_2 \tilde g_3 \tilde g_5$ & $0.7036 \pm 0.0066$ &  &  &  \\
$\tilde g_2 \tilde g_3 \tilde g_6$ & $0.7468 \pm 0.0056$ &  &  &  \\
$\tilde g_2 \tilde g_4 \tilde g_5$ & $0.7038 \pm 0.0066$ &  &  &  \\
$\tilde g_2 \tilde g_4 \tilde g_6$ & $0.8285 \pm 0.0062$ &  &  &  \\
$\tilde g_2 \tilde g_5 \tilde g_6$ & $0.6861 \pm 0.0058$ &  &  &  \\
$\tilde g_3 \tilde g_4 \tilde g_5$ & $0.7357 \pm 0.0083$ &  &  &  \\
$\tilde g_3 \tilde g_4 \tilde g_6$ & $0.7484 \pm 0.0056$ &  &  &  \\
$\tilde g_3 \tilde g_5 \tilde g_6$ & $0.6625 \pm 0.0051$ &  &  & \checkmark \\
$\tilde g_4 \tilde g_5 \tilde g_6$ & $0.6394 \pm 0.0060$ &  &  &  \\
$\tilde g_1 \tilde g_2 \tilde g_3 \tilde g_4$ & $0.6067 \pm 0.0074$ &  &  &  \\
$\tilde g_1 \tilde g_2 \tilde g_3 \tilde g_5$ & $0.5391 \pm 0.0086$ &  &  &  \\
$\tilde g_1 \tilde g_2 \tilde g_3 \tilde g_6$ & $0.4334 \pm 0.0063$ & \checkmark &  &  \\
$\tilde g_1 \tilde g_2 \tilde g_4 \tilde g_5$ & $0.4247 \pm 0.0093$ &  &  &  \\
$\tilde g_1 \tilde g_2 \tilde g_4 \tilde g_6$ & $0.3960 \pm 0.0077$ & \checkmark &  &  \\
$\tilde g_1 \tilde g_2 \tilde g_5 \tilde g_6$ & $0.4435 \pm 0.0076$ & \checkmark &  &  \\
$\tilde g_1 \tilde g_3 \tilde g_4 \tilde g_5$ & $0.5897 \pm 0.0074$ &  &  &  \\
$\tilde g_1 \tilde g_3 \tilde g_4 \tilde g_6$ & $0.4349 \pm 0.0080$ & \checkmark &  &  \\
$\tilde g_1 \tilde g_3 \tilde g_5 \tilde g_6$ & $0.4465 \pm 0.0061$ & \checkmark &  &  \\
$\tilde g_1 \tilde g_4 \tilde g_5 \tilde g_6$ & $0.4465 \pm 0.0061$ & \checkmark &  &  \\
$\tilde g_2 \tilde g_3 \tilde g_4 \tilde g_5$ & $0.7037 \pm 0.0066$ &  &  &  \\
$\tilde g_2 \tilde g_3 \tilde g_4 \tilde g_6$ & $0.7465 \pm 0.0056$ &  &  &  \\
$\tilde g_2 \tilde g_3 \tilde g_5 \tilde g_6$ & $0.6113 \pm 0.0063$ &  &  &  \\
$\tilde g_2 \tilde g_4 \tilde g_5 \tilde g_6$ & $0.6860 \pm 0.0058$ &  &  &  \\
$\tilde g_3 \tilde g_4 \tilde g_5 \tilde g_6$ & $0.6624 \pm 0.0051$ &  &  &  \\
$\tilde g_1 \tilde g_2 \tilde g_3 \tilde g_4 \tilde g_5$ & $0.4235 \pm 0.0093$ &  &  &  \\
$\tilde g_1 \tilde g_2 \tilde g_3 \tilde g_4 \tilde g_6$ & $0.3735 \pm 0.0078$ & \checkmark &  &  \\
$\tilde g_1 \tilde g_2 \tilde g_3 \tilde g_5 \tilde g_6$ & $0.4071 \pm 0.0077$ & \checkmark &  &  \\
$\tilde g_1 \tilde g_2 \tilde g_4 \tilde g_5 \tilde g_6$ & $0.5059 \pm 0.0052$ & \checkmark &  &  \\
$\tilde g_1 \tilde g_3 \tilde g_4 \tilde g_5 \tilde g_6$ & $0.4884 \pm 0.0057$ & \checkmark &  &  \\
$\tilde g_2 \tilde g_3 \tilde g_4 \tilde g_5 \tilde g_6$ & $0.6112 \pm 0.0063$ &  &  &  \\
$\tilde g_1 \tilde g_2 \tilde g_3 \tilde g_4 \tilde g_5 \tilde g_6$ & $0.4046 \pm 0.0060$ & \checkmark &  &  \\
\hline \hline
\end{tabular}
 \end{table*}


\end{document}